\documentclass[epj]{svjour}
\usepackage{graphics}
\usepackage{psfrag}
\usepackage{times}
\usepackage{amsmath, amssymb}
\usepackage{subfigure,rotating}

\newcommand{\beq}{\begin{equation}}
\newcommand{\eeq}{\end{equation}}
\newcommand{\beqa}{\begin{eqnarray}}
\newcommand{\eeqa}{\end{eqnarray}}

\begin{document}
\renewcommand{\arraystretch}{1.2}
\title{Coulomb effects in polarization transfer in elastic antiproton
and proton electron scattering at low energies.} 

\author{H. Arenh\"ovel}

\institute{
  Institut f\"ur Kernphysik, Johannes Gutenberg-Universit\"at Mainz,
  D-55099 Mainz, Germany}

\date{Received: \today / Revised version: date}

\abstract{The influence of Coulomb distortion on the polarization
  transfer in elastic proton and antiproton electron scattering at low
  energies is calculated in a distorted wave Born approximation. For
  antiproton electron scattering Coulomb effects reduce substantially
  the polarization transfer cross section compared to the plane wave Born
  approximation whereas for proton electron scattering they lead to a
  dramatic increase for kinetic proton lab energies below about 20 keV.}

\PACS{{13.88.+e}{Polarization in interactions and scattering} \and 
{25.30.Bf}{Elastic electron scattering} \and {29.27.Hj}{Polarized beams} }
   
\maketitle

\section{Introduction                                            \label{intro}}
Spin degrees of freedom constitute an important ingredient of the
internal dynamics of elementary particles and thus the study of polarization
observables provides important information on their 
intrinsic properties. In order to investigate such polarization observables
experimentally, one needs sufficiently intensive sour\-ces of polarized
beams on the one hand and efficient polarimeters on the other hand. Indeed,
significant improvements of experimental methods for polarizing
particles and their analysis over the last decades have made possible
a great variety of different experiments devoted to the study of polarization
observables. 

One interesting method for polarizing particles is the use of polarization
transfer in a storage ring, where a beam of easily polarized particles
like, e.g.\ electrons, scatters elastically with polarization transfer on
unpolarized particles, e.g.\ antiprotons. Although the probability of
such a polarization transfer in one scattering event might be small, a series of
successive scatterings in the storage ring can lead to a significant
polarization build-up for the originally unpolarized particles~\cite{HoM94,WaA07}. 
The case of the antiproton is particularly interesting because sources
for polarized antiprotons are still lacking. Such a source would allow
for studies of iso-spin and spin symmetries in the interaction of
nucleons. 

In principle the polarization transfer from polarized electrons to unpolarized
nucleons has been
calculated in lowest order in the framework of QED a long time
ago~\cite{Sco59,Sco66,Dom69,Arn81,ArL88} and extensively used for the
measurement of the electric form factors of neutron and proton (see
e.g.~\cite{Glazier:2004ny,Perdrisat:2003xj}). However, these
calculations are badly adapted to the situation of the interaction of
electron and antiproton beams in a storage ring at small
relative kinetic energies, because Coulomb effects cannot
be neglected at such low energies in contrast to the above cited
applications at higher energies. This has been pointed out by Horowitz
and Meyer~\cite{HoM94}, estimating the influence of Coulomb effects in
a rough approximation. 

The present work aims at a more reliable treatment of Cou\-lomb effects
on the polarization transfer cross section based on a distorted wave
approximation.

\section{The polarization transfer cross section   \label{crossection}}

For the scattering of an unpolarized spin-one-half hadron (here proton
or antiproton) on a polarized electron, the general
expression for the polarization component $P^h_k$ of the outgoing
hadron is given in the centre-of-momentum (c.m.) frame by 
\beqa
P^h_k\,\frac{d\sigma(S_e(l))}{d\Omega_h}&=&
\frac{1}{8\pi^2}\,\frac{M_e^2M_h^2}{W^2}\,\nonumber\\
&&\times \mbox{Trace}(T^\dagger_{fi}
\rho_f^h(S_h(k)) T_{fi}\rho_i^e(S_e(l)))\,,\label{pol-cross}
\eeqa
where the initial electron, polarized along an axis ``$l$'', is
described by the density matrix 
\beq
\rho_i^e(S_e(l))=\frac{1}{2}(1+\gamma_5{S}\:\!\!\!\!\!/_e(l))\,,
\label{e-density}
\eeq
and the outgoing hadron polarization along an axis ``$k$'' by
\beq
\rho_f^h(S_h(k))=\gamma_5{S}\:\!\!\!\!\!/_h(k)\,,
\eeq
with the relativistic spin vectors $S_e(l)$ and $S_h(k)$ of
electron and hadron, respectively, 
\beqa
S_e(l)&=& \Big(-\frac{1}{M_e}\,\hat{\vec s}(l)\cdot \vec p,\,
\hat{\vec s}(l)+ \frac{\hat{\vec s}(l)\cdot \vec
p}{M_e(E_e+M_e)}\,\vec p\Big),\\  
S_h(k)&=& \Big(\frac{1}{M_h}\,\hat{\vec s}(k)\cdot \vec p',\,
\hat{\vec s}(k)+ \frac{\hat{\vec s}(k)\cdot \vec
p'}{M_h(E_h'+M_h)}\,\vec p'\Big). 
\eeqa
Here $\hat{\vec s}(k)$ denotes a unit vector pointing into the
direction of the $k$-axis. The trace in (\ref{pol-cross}) is to be
taken over the hadron and electron spin degrees of freedom, and
$T_{fi}$ denotes the scattering matrix. Furthermore, $W=E_h+E_e$
denotes the invariant energy of the hadron-electron system. Initial
and final hadron and electron momenta are denoted 
by 
\beq
p_h^{(\prime)}=(E_h^{(\prime)},\vec p^{\,(\prime)})\quad\mbox{and}
\quad p_e^{(\prime)}=(E_e^{(\prime)},-\vec p^{\,(\prime)})\,,
\eeq
and their masses by $M_h$ and $M_e$, respectively. 

The second contribution of the electron density matrix in
(\ref{e-density}) describes the polarization transfer $P_{kl}$ from an
incoming electron polarized along the $l$-axis to the outgoing hadron
polarized a\-long the $k$-axis, i.e.\
\beqa
P_{kl}\,\frac{d\sigma^0}{d\Omega_h}&=&
\frac{1}{16\pi^2}\,\frac{M_e^2M_h^2}{W^2}\,\nonumber\\
&&\times \mbox{Trace}(T^\dagger_{fi}\gamma_5{S}\:\!\!\!\!\!/_h(k)  
T_{fi}\gamma_5{S}\:\!\!\!\!\!/_e(l))\,,\label{pol_trans}
\eeqa
where ${d\sigma^0}/{d\Omega_h}$ denotes the unpolarized scattering
cross section in the c.m.\ system
\beqa
\frac{d\sigma^0}{d\Omega_h}&=&
\frac{1}{16\pi^2}\,\frac{M_e^2M_h^2}{W^2}\,
\mbox{Trace}(T^\dagger_{fi}T_{fi})\,.
\label{unpol_xsection}
\eeqa

In the c.m.\ 
frame we use as reference system the $z$-axis along the incoming 
hadron momentum $\vec p$. Since we are mainly interested in the integrated
polarization transfer cross section, we can fix the $y$-axis to an
arbitrary direction in the plane perpendicular to the $z$-axis. Then
the $x$-axis is chosen to form a right handed orthogonal system. 

We will discuss three different approximations for the scattering
matrix, namely (i) the plane wave Born approximation (PW) and for the
inclusion of Coulomb effects (ii) the approximation of Horowitz and
Meyer (HM)~\cite{HoM94}, and (iii) the distorted wave Born
approximation (DW). 

\subsection{Plane wave approximation                      \label{spwba}}
The evaluation of the lowest order Feynman diagram, the
one-photon-exchange approximation, which corresponds to the PW,
yields for the scattering matrix, indicating only the initial and
final spin projections $m_{e/h}^{(\prime)}$ with respect to a
given quantization axis, 
\beq
T_{m_h' m_e',\,m_h m_e}=\langle
p_h'\,m_h'|j_h^\mu|p_h\,m_h\rangle  
\frac{1}{q_\nu^2}\langle
p_e'\,m_e'|j_{e,\,\mu}|p_h\,m_e\rangle \,,\label{pwba} 
\eeq
with $q_\mu=p_{h,\mu}-p_{h,\mu}'$ as four momentum transfer, and
denoting the electron 
and hadron current operators by $j_{e,\,\mu}$ and $j_{h,\,\mu}$ and
their Dirac spinors by $|p_e\,m_e\rangle$ and $|p_h\,m_h\rangle$,
respectively. 

Evaluation of the trace leads to the following expression of the
polarization transfer cross section in the c.m.\ system~\cite{Sco59}
\beqa
P_{kl}\frac{d\sigma^0}{d\Omega_h}&=&\frac{4\alpha^2}{Q^4}\,\frac{M_eM_h}
{W^2}\,\nonumber\\
&&\hspace*{-1cm}\times G_M\Big[G_E(S_e(l)\cdot q\, S_h(k)\cdot
q-q_\mu^2 S_e(l)\cdot S_h(k))\nonumber\\&&\hspace*{-1cm}
-\eta\tau(G_M-G_E)S_e(l)\cdot q\, S_h(k)\cdot (p_h+p_h')\Big]\,,
\label{spintransfer}
\eeqa
and the well-known unpolarized c.m.\ scattering cross section 
\beqa
\frac{d\sigma^0}{d\Omega_h}&=&\frac{4\alpha^2p^2}{Q^4}\,
\Big[\tau(G_E^2+\eta G_M^2)
\Big(\cos^2(\theta/2) +\frac{M_e^2M_h^2}{p^2W^2\tau}\,\Big)\nonumber\\
&&+2\frac{M_h^2}{W^2}\,\Big(\eta-\frac{M_e^2}{2M_h^2}\Big)G_M^2
\sin^2(\theta/2)\Big]\,,
\label{xsection}
\eeqa
where $\alpha$ denotes the fine structure constant. Furthermore, $G_E$
and $G_M$ stand for the electric and magnetic hadron Sachs form
factors, respectively, and ($Q^2=-q_\mu^2=\vec q^2$) 
\beq
\tau=(1+\eta)^{-1}\,\mbox{ with }\,\, \eta=\frac{Q^2}{4\,M_h^2}\,.
\eeq
One should note that (\ref{xsection}) is not the usual high energy
limit ($M_e\to 0$). 

For low energy hadrons ($p^2/2M_h\ll M_h$) we can adopt non-relativistic
kinematics and then the last term in Eq.\ (\ref{spintransfer}) can
safely be neglected. Furthermore, $G_E$ can be replaced by the hadron
charge $Z_h$ and $G_M$ by $Z_h\mu_h$, the hadron magnetic moment in
units of $e/2M_h$. The resulting expression is the one Horowitz and
Mayer~\cite{HoM94} have used. 

In addition, we will briefly consider for later purposes the
low-energy expansion of the scattering matrix in
(\ref{pwba}). Expanding the currents up to order $(p/M_{e/h})^2$, one
finds in the c.m.-frame
\beq
T_{m_h' m_e',\,m_h m_e}=
\chi_{m_h'}^\dagger\chi_{m_e'}^\dagger \widetilde T^{nr}
\chi_{m_e}\chi_{m_h}\,,
\eeq
where $\chi_{m_{e/h}}$ denotes an electron or hadron Pauli
spinor, respectively, and
\beqa
\widetilde T^{nr}&=&-\frac{4\pi\alpha Z_h}{\vec q^2}
\Big[1-\Big(\frac{1}{8M_e^2}+\frac{2\mu_h-1}{8M_h^2}\Big)\vec q^2
+\frac{{\vec P}^2}{4M_eM_h}\nonumber\\ 
&&-\Big(\frac{1}{8M_e^2}+\frac{1}{4M_eM_h}\Big) 
i(\vec\sigma_e\times\vec q)\cdot\vec P)\nonumber\\ 
&&-\Big(\frac{\mu_h}{4M_eM_h}-\frac{2\mu_h-1}{8M_h^2}\Big)
i(\vec\sigma_h\times\vec q)\cdot\vec P\nonumber\\ 
&& -\frac{\mu_h}{4M_eM_h}(\vec q^2\vec\sigma_e\cdot\vec\sigma_h
-\vec\sigma_e\cdot\vec q \vec\sigma_h\cdot\vec q)\Big]
\eeqa
with $\vec P=\vec p+\vec p'$. Keeping only the very lowest
order term for each of the spin-independent and the linear and
quadratic spin dependent parts, one finds
finally 
\beqa
T^{nr}&=&-\frac{4\pi\alpha Z_h}{\vec q^2}\Big[1-\frac{i}{4M_e^2}
(\vec\sigma_e\times\vec q)\cdot\vec p\nonumber\\ 
&& -\frac{\mu_h}{4M_eM_h}(\vec q^2\vec\sigma_e\cdot\vec\sigma_h
-\vec\sigma_e\cdot\vec q \vec\sigma_h\cdot\vec q)\Big]\,,\label{Tpw_nr}
\eeqa
where $\vec q\times\vec P=2\vec q\times\vec p$ has been used.
The second term describes the electron spin-orbit interaction and the
last one the hyperfine interaction between electron and
hadron. Furthermore, for the polarization vector one has as
nonrelativistic limit
\beq
\langle pm'|\gamma_5{S}\:\!\!\!\!\!/_e(l)|pm\rangle\to
\chi_{m'}^\dagger\sigma_l\chi_m 
\eeq
and thus the nonrelativistic limit of (\ref{pol_trans})
becomes
\beqa
P_{kl}\,\frac{d\sigma^0}{d\Omega_h}&=&
\frac{1}{16\pi^2}\,\frac{M_e^2M_h^2}{W^2}
\mbox{Trace}(T^{nr\,\dagger}_{fi}\sigma_kT^{nr}_{fi}\sigma_l)\,\nonumber\\
&=&\frac{4c_s}{q^2}\,(\widehat q_k\widehat q_l-\delta_{kl})\,,
\label{pol_trans_nr}
\eeqa
where $\widehat{\vec q}$ denotes the unit vector along $\vec q$ and
\beq
c_s=\frac{\alpha^2Z_h^2\mu_h M_eM_h}{W^2}\,.
\eeq 

\subsection{Coulomb effects  \label{coulomb}}
As already mentioned, at those low energies considered here, Coulomb
effects cannot be neglected~\cite{HoM94}. In order to incorporate them,
we adopt a non-relativistic framework with the Coulomb potential $V^C$
as the main interaction between hadron and electron. Since we want to
describe polarization transfer effects, we have to take into account spin
degrees of freedom, which means we have to include in addition lowest order
relativistic contributions to the electromagnetic interaction between
hadron and electron. Thus the total interaction consists of the spin
independent static Coulomb potential $V^C$, the spin-orbit interaction
$V^{LS}_e$ of the electron and the hyperfine interaction $V^{SS}$,
neglecting the much smaller hadron spin-orbit interaction, i.e.\
\beq
V=V^C+V^{LS}_e+V^{SS}=V^C+V^\sigma\,,\label{potential}
\eeq
with
\beqa
V^C&=& -\frac{4\pi\alpha Z_h}{r}\,,\\
V^{LS}_e&=&c^{LS}_e(\vec\nabla V^C\times\vec
p\,)\cdot\vec\sigma_e)\,,\\
V^{SS}&=&c^{SS}\Big(\vec\sigma_e\cdot\vec\sigma_h\Delta-
(\vec\sigma_e\cdot\vec\nabla)(\vec\sigma_h\cdot\vec\nabla)\Big)V^C\,.
\label{int_pot}
\eeqa
Here $\vec r$ and $\vec p$ denote relative coordinate and momentum,
and the constants are
\beq
c^{LS}_e=\frac{1}{4M_{e}^2}\quad\mbox{and}\quad
c^{SS}=\frac{\mu_h}{4M_eM_h}\,.
\eeq
It is worthwhile to point out that the plane wave Born approximation
of the scattering matrix for the potential in (\ref{potential}) is
exactly the low energy expansion in (\ref{Tpw_nr}).

The complete scattering matrix for the interaction $V$
\beq
T_{fi}=\langle\phi_f|T|\phi_i\rangle\,,
\eeq
with $|\phi_{i/f}\rangle$ as plane waves, can be expressed in terms of
the scattering matrix $T^C$ of the pure Coulomb potential $V^C$ plus an
additional contribution from the spin interaction $T^\sigma$
\beq
T=T^C+T^\sigma\,,\label{t_tilde}
\eeq
where 
\beq
T^\sigma=(1+T^CG_{(+)}^0)\widetilde
T^\sigma(1+G_{(+)}^0T^C)\,,\label{tsigma} 
\eeq
and the auxiliary scattering matrix $\widetilde T^\sigma$ is generated
by the spin interaction $V^\sigma$ and determined as solution of
\beq
\widetilde T^\sigma=V^\sigma(1+G_{(+)}^C\widetilde T^\sigma)\,.
\eeq
Here, $G_{(+)}^0$ denotes the free propagator and $G_{(+)}^C$ the
propagator in the Coulomb field 
\beq
G_{(+)}^C=G_{(+)}^0(1+T^CG_{(+)}^0)\,.
\eeq
With the help of the representation (\ref{t_tilde}) one obtains
\beqa
T_{fi}&=&\langle\phi_f|T^C+(1+T^CG_{(+)}^0)\widetilde T^\sigma(1+G_{(+)}^0T^C)
|\phi_i\rangle\nonumber\\
&=&T_{fi}^C+\langle \psi_{f}^{C(-)}|\widetilde T^\sigma
|\psi_{i}^{C(+)}\rangle\,.\label{DW}
\eeqa
The nonrelativistic Coulomb amplitude is given by~\cite{Mes69}
\beq
T_{fi}^C=-\frac{4\pi \alpha Z_h}{{\vec q}^2}\,
e^{i\phi_c(\theta)}\,,\label{t_c}
\eeq
where $\phi_c$ denotes the Coulomb phase 
\beqa
\phi_c(\theta)&=&-\eta_c \ln(\sin^2(\theta/2))+2\sigma_c\,,\\
\sigma_c&=&\arg(\Gamma(1+i\eta_c))\,,
\eeqa
with $\Gamma(z)$ standing for the gamma function and the Sommerfeld
Coulomb parameter $\eta_c=-\alpha Z_h/v$, where $v=p/M$ denotes the
relative velocity between hadron and electron and $M=M_eM_h/(M_e+M_h)$
the reduced electron-hadron mass. 

Furthermore, in (\ref{DW}) $\psi_{i/f}^{C(\pm)}$ represent incoming
and outgoing scattering solutions for the pure Coulomb field, and we
have used the relation 
\beq
|\psi_{i}^{C(+)}\rangle=(1+G_{(+)}^0T^C)|\phi_i\rangle\,
\eeq
for the incoming scattering wave and a corresponding one for the
outgoing wave $|\psi_{f}^{C(-)}\rangle$. Explicit analytic expressions
for these Cou\-lomb scat\-tering waves are well known, e.g.~\cite{Mes69}, 
\beq
\psi^{C(+)}_{\vec p}(\vec r)=N(\eta_c)e^{i\vec p\cdot\vec r}
\,_1F_1(-i\eta_c,1;i(pr-\vec p\cdot\vec r\,))
\eeq
and
\beq
\psi^{C(-)}_{\vec p}(\vec r\,)=(\psi^{C(+)}_{-\vec p}(\vec r\,))^*\,,
\eeq
where $_1F_1(a,b;z)$ denotes the confluent hypergeometric
function, and the normalization factor is given by
\beqa
N(\eta_c)&=&e^{-\frac{\pi}{2}\eta_c}\Gamma(1+i\eta_c)\nonumber\\
&=&\sqrt{\frac{2\pi\eta_c}{e^{2\pi\eta_c}-1}}\,e^{i\sigma_c}
\,.
\eeqa

In lowest order (PW), $T^\sigma$ is given by the spin interaction
$V^\sigma$ alone, whereas the distorted wave Born approximation (DW) is
defined by replacing in (\ref{tsigma}) $\widetilde T^\sigma$ by
$V^\sigma$ yielding  
\beq
T_{fi}^{DW}=T_{fi}^C+\langle
\psi_{f}^{C(-)}|V^\sigma|\psi_{i}^{C(+)}\rangle\,,
\eeq
which means in comparison to the PW the replacement of the plane
waves by Cou\-lomb distor\-ted scat\-tering waves. In \cite{HoM94}
Cou\-lomb effects were included in a rough approximation by using the
exact non-relativistic Coulomb amplitude in (\ref{t_c}) for the
spin-in\-depen\-dent part while taking the hyper\-fine am\-plitude in
PW multiplied solely by the Coulomb wave functions at the origin,
i.e.\ by $N(\eta_c)^2$. In the present work, we have evaluated the
hyperfine interaction in the full DW whereas we have neglected the
spin-orbit interaction $V^{LS}_e$ because it will not contribute to the
polarization-transfer in lowest order proportional to $c^{SS}$, since 
$V^{LS}_e$ is linear in the electron spin variables and thus contributes 
only via the interference between $V^{LS}_e$ and $V^{SS}$ which we 
consider as higher order. 

According to the structure of $V^{SS}$, consisting of a scalar and
a traceless tensor part, the $T$ matrix has a corresponding general
form with respect to the spin degrees of freedom 
\beq
T=-4\pi\alpha Z_h\Big(\frac{a}{q^{2}} -c^{SS}\,(d\,\vec \sigma_e\cdot
\vec \sigma_h 
+\vec \sigma_e\cdot\vec{D}\cdot\vec \sigma_h)\Big)\,,\label{T_general}
\eeq
where the tensor $\vec{D}$ of rank two is symmetric and
traceless. This tensor as well as the scalars $a$ and $d$ depend on
the type of approximation. In detail one has with $q^2=\vec q^{2}$
\begin{enumerate}
\item[(i)]
PW:
\beqa
a^{PW}&=&1\,,\\
d^{PW}&=&\frac{2}{3}\,,\label{s_born}\\
{D}_{ij}^{PW}&=&-(\widehat q_i\widehat q_j-\frac{1}{3}
\delta_{ij})\,.\label{t_born}
\eeqa
\item[(ii)]
Horowitz-Meyer (HM):
\beqa
a^{HM}&=&e^{i\phi_c(\theta)}\,,\\
d^{HM}&=&N(\eta_c)^2d^{PW}\,,\label{s_hm}\\
{D}_{ij}^{HM}&=&N(\eta_c)^2D_{ij}^{PW}\,.\label{t_hm}
\eeqa
\item[(iii)]
DW:
\beqa
a^{DW}&=&e^{i\phi_c(\theta)}=a^{HM}\,,\\
d^{DW}&=&d^{HM}\label{s_dwba}\,,\\
{D}_{ij}^{DW}&=&\frac{1}{4\pi}\int \frac{d^3r}{r^3}
 \psi^{C(-)}_{\vec p^{\,\prime}}(\vec r)^*\nonumber\\
&&(3\hat{r}_i\,\hat{r}_j-\delta_{ij})\psi^{C(+)}_{\vec p}(\vec
r)\label{dij} \,.\label{t_dwba}
\eeqa
\end{enumerate}

\subsubsection{The unpolarized cross section}

The spin dependent part of the scattering matrix gives also a
contribution to the unpolarized differential cross section because one
finds 
\beqa
\mbox{Trace}(T^\dagger_{fi}T_{fi})&=&(8\pi\alpha Z_h)^2
\Big(\frac{|a|^2}{q^4}\nonumber\\
&&+(c^{SS})^2({3}|d|^2  +\sum_{ij}|D_{ij}|^2)\Big),
\eeqa
and, therefore, for the unpolarized non\-relati\-vistic cross section
\beqa 
\frac{d\sigma^0}{d\Omega_h}
&=&\frac{d\sigma_R}{d\Omega_h}\Big(1+\frac{q^4}{|a|^2}
(c^{SS})^2({3}|d|^2 
+\sum_{ij}|D_{ij}|^2)\Big)\,,
\eeqa
where ${d\sigma_R}/{d\Omega_h}=4\alpha^2Z_h^2M_e^2/q^4$ denotes the
un\-polarized non\-relati\-vistic Ruther\-ford cross section, and the
term proportional to $(c^{SS})^2$ describes the relative contribution
from the hyperfine interaction. Its size can serve as a criterion
for the validity of the DW approximation, i.e.\ as long as its size is
small compared to one the approximation should work well.

\subsubsection{The polarization transfer cross section}
For the polarization transfer cross section one has to evaluate 
(\ref{pol_trans}) with the $T$-matrix of (\ref{T_general}) and obtains, 
neglecting terms proportional to $(c^{SS})^2$ as higher order,
\beqa
\mbox{Trace}(T^\dagger_{fi}\sigma^e_lT_{fi}\sigma^h_k)&=&
-\frac{c^{SS}}{q^{2}}(8\pi\alpha Z_h)^2\widehat S_{kl}\,,
\eeqa
where the second-rank tensor $\widehat S_{ij}$ is defined by
\beqa
\widehat S_{ij}&=&\Re e[a^*(d\,\delta_{ij}+D_{ij})]\nonumber\\
&=&S_0\delta_{ij}+S_{ij}\,.
\eeqa
One should note that $S_{ij}$ is symmetric and traceless.
According to the approximations in (\ref{t_born}) through
(\ref{t_dwba}) one finds
\begin{enumerate}
\item[(i)]
PW:
\beqa
S_0^{PW}&=&\frac{2}{3}\,,\\
S_{ij}^{PW}&=&-(\widehat q_i\widehat q_j -\frac{1}{3} \delta_{ij})\,. 
\eeqa
\item[(ii)]
Horowitz-Meyer (HM):
\beqa
S_0^{HM}&=&\frac{2}{3}\frac{2\pi\eta_c}{e^{2\pi\eta_c}-1}\cos[\eta_c
\ln(\sin^2(\theta/2))]\,,\\
S^{HM}_{ij}&=&\frac{2\pi\eta_c}{e^{2\pi\eta_c}-1}\cos[\eta_c
\ln(\sin^2(\theta/2))]S^{PW}_{ij}\,. 
\eeqa
\item[(iii)]
DW:
\beqa
S_0^{DW}&=&S_0^{HM}\,,\\
S^{DW}_{ij}&=&\frac{\eta_c}{2(e^{2\pi\eta_c}-1)}\,
\Re e \Big[ e^{i\eta_c\ln(\sin^2(\theta/2))}D_{ij}^{DW}\Big].
\eeqa
\end{enumerate}

The resulting nonrelativistic polarization transfer cross section
corresponding to (\ref{spintransfer}) then reads in the c.m.\ system 
\beq
P_{ij}\frac{d\sigma^0}{d\Omega_h}=-\frac{\alpha^2Z_h^2\mu_hM_eM_h}{q^2W^2}
\,\widehat S_{ij}\,. 
\label{spintransfer_C}
\eeq
We would like to point out that the nonrelativistic limit of
(\ref{spintransfer}), given in (\ref{pol_trans_nr}), coincides with
the PW of (\ref{spintransfer_C}). 

The polarization transfer tensor $\widehat S_{ij}$ depends on the scattering
angle 
$\Omega_h=(\theta,\phi)$, i.e.\ $\widehat S_{ij}(\theta,\phi)$. However, it
suffices to calculate $\widehat S_{ij}$ for $\phi=0$, from which one obtains
$\widehat S_{ij}(\theta,\phi)$ for an arbitrary $\phi$ by a rotation around the
$z$-axis by an angle $\phi$ 
\beq
\widehat S_{ij}(\theta,\phi)=\Big(R(\phi)\widehat
S^0(\theta)R^{-1}(\phi)\Big)_{ij}\,, 
\eeq
defining 
\beq
\widehat S^0_{ij}(\theta)=\widehat S_{ij}(\theta,0)
=S^0_0(\theta)\delta_{ij}+S^0_{ij}(\theta)\,,
\eeq
and the rotation matrix 
\beq
R(\phi)=\left(\begin{matrix}
\cos\phi&-\sin\phi&0\\
\sin\phi&\cos\phi&0\\
0&0&1
\end{matrix}\right)\,.
\eeq
Since $\widehat S_{ij}$ is symmetric and $\widehat S^0_{xy}=\widehat
S^0_{zy}=0$ one finds as $\phi$-dependence of the tensor components
\beqa
\widehat S_{xx}(\theta,\phi)&=&\widehat S^0_{xx}(\theta)\cos^2\phi
+\widehat S^0_{yy}(\theta)\sin^2\phi\nonumber\\
&=&S_0^0+ S^0_{xx}(\theta)\cos^2\phi
+ S^0_{yy}(\theta)\sin^2\phi\,,\label{phi_xx}\\
\widehat S_{yy}(\theta,\phi)&=&\widehat S^0_{xx}(\theta)\sin^2\phi
+\widehat S^0_{yy}(\theta)\cos^2\phi\nonumber\\
&=&S_0^0+ S^0_{xx}(\theta)\sin^2\phi
+ S^0_{yy}(\theta)\cos^2\phi\,,\\
\widehat S_{zz}(\theta,\phi)&=&S_0^0+ S^0_{zz}(\theta)\,,\\
\widehat S_{xy}(\theta,\phi)&=&(S^0_{xx}(\theta)-S^0_{yy}
(\theta))\cos\phi\sin\phi\,,\\
\widehat S_{xz}(\theta,\phi)&=& S^0_{xz}(\theta)\cos\phi\,,\\
\widehat S_{yz}(\theta,\phi)&=& S^0_{xz}(\theta)\sin\phi\,.\label{phi_yz}
\eeqa

\section{Results for the differential and integrated polarization transfer
cross sections} \label{crossintegr}
As it turns out in the explicit evaluation, the tensor $D_{ij}$,
defined in (\ref{t_dwba}), contains the dominant 
Cou\-lomb effect while the scalar part $d$ is negligibly small. In
principle, one could calculate 
numerically the three-dimensional integral. However, in view of the
strongly oscillating Coulomb wave functions and the slow convergence
of the integral with $r\to \infty$, it is more advantageous to use an
integral representation recently proposed by Levin, Alt, and
Yakovlev~\cite{LeA01}. With the
help of this integral representation the space integral can be
performed analytically leaving a two-dimensional integral, which is
easier to evaluate numerically (details may be found in the appendix). 

\begin{figure}[ht]
\includegraphics[width=1.\columnwidth]{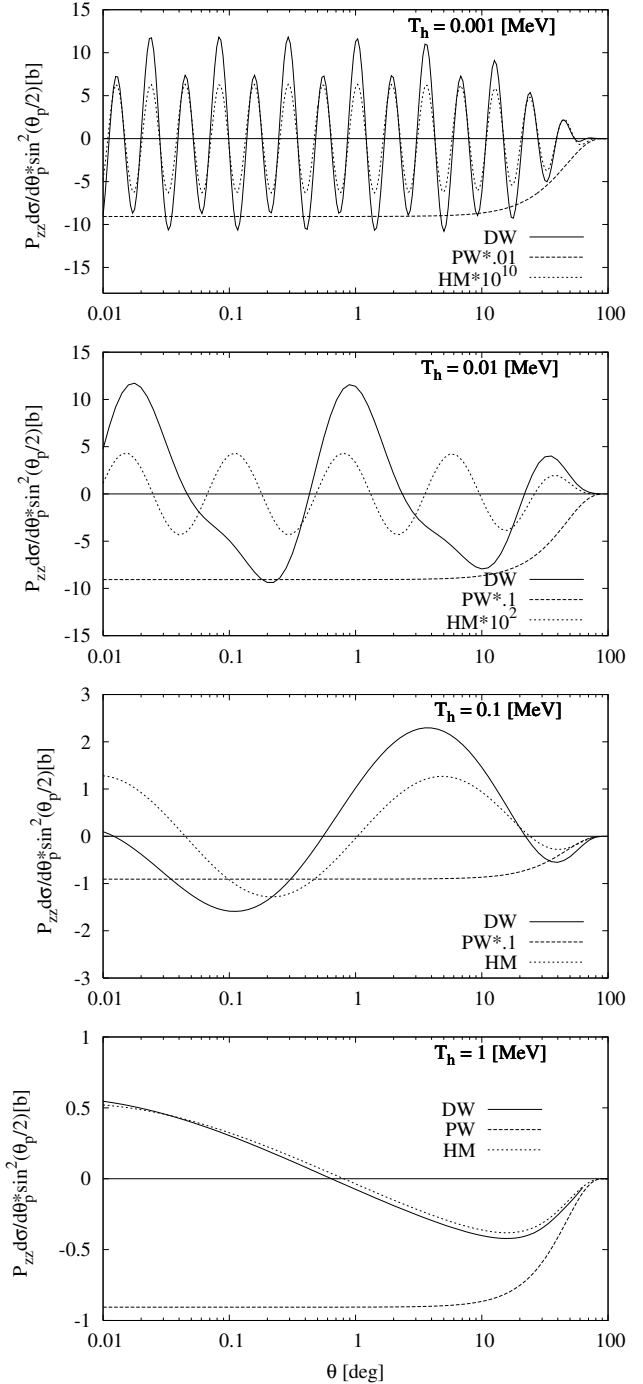}
\caption{The differential polarisation transfer cross section weighted
by $\sin^2(\theta/2)$ at $\phi=0$ for electron antiproton scattering
in the c.m.\ frame for various antiproton lab kinetic energies and for
the approximations PW, HM, and DW.}
\label{diffsigA} 
\end{figure}

\begin{figure}[ht]
\includegraphics[width=1.\columnwidth]{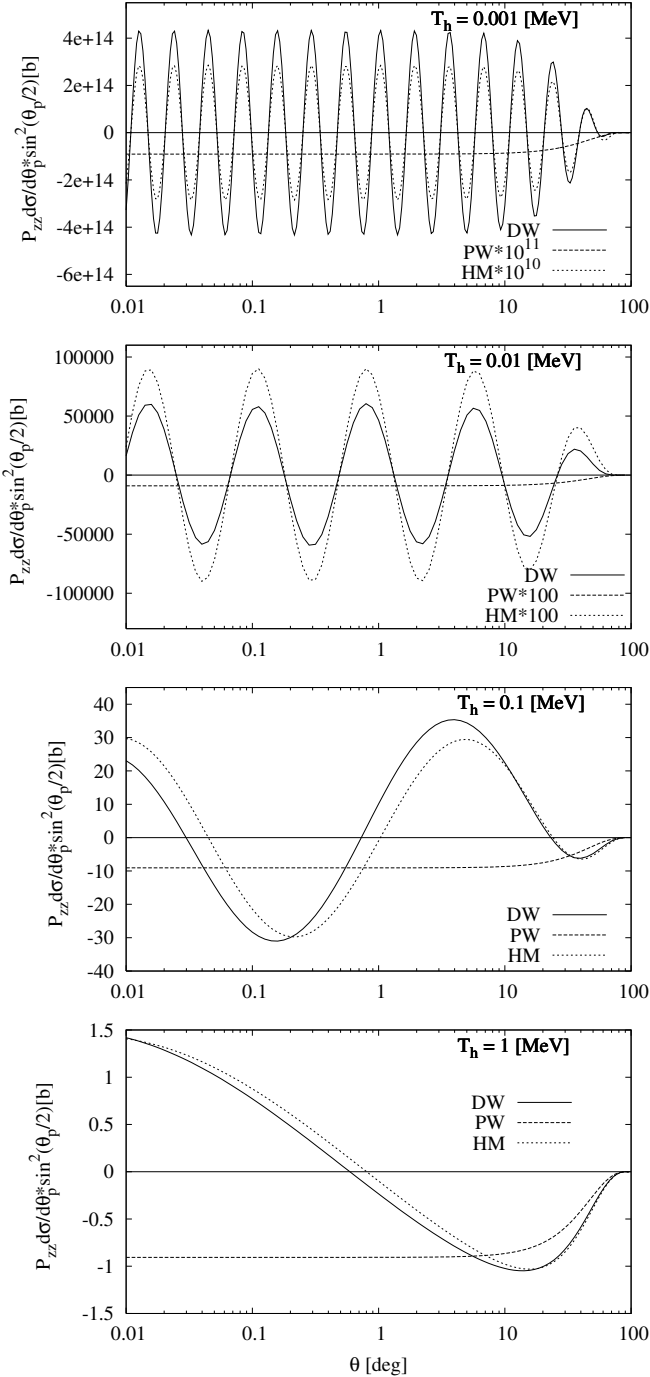}
\caption{The differential polarisation transfer cross section weighted
by $\sin^2(\theta/2)$ at $\phi=0$ for electron proton scattering in
the c.m.\ frame for various proton lab kinetic energies and for
the approximations PW, HM, and DW.}
\label{diffsigP} 
\end{figure}

We begin with presenting first the differential polarization transfer cross
sections $P_{ij}{d\sigma^0}/{d\Omega_h}$ at $\phi=0$ for
antiproton electron scattering in Fig.~\ref{diffsigA} and 
for proton electron scattering in Fig.~\ref{diffsigP} for hadron
kinetic lab energies between 0.001 and 1~MeV for the three
approximations ``PW'', ``HM'' and ``DW''. The cross sections have been
weighted by $\sin^2(\theta/2)$ in order to account for their
$1/q^2$-dependence. 

While in PW the angular behaviour weighted
by $\sin^2(\theta/2)$ is very smooth
and approaching a constant value for $\theta\to 0$, Coulomb effects
introduce an increasing oscillatory behaviour with decreasing hadron
lab energy with an almost constant amplitude for $\theta\to
0$. Qualitatively the Coulomb effects look quite similar for
antiproton and proton electron scattering, except for the size. The
reason for this similarity is the symmetry property displayed in
eqs.~(\ref{symreal}) and (\ref{symimag}) of the appendix, whereas the
size is governed 
by the normalization factor $N(\eta)^2$ in (\ref{sym}) reflecting the
fact, that for antiprotons the Coulomb field acts repulsive and thus
produces a decrease of the scattering wave near the origin while for 
protons it acts attractive and leads to a strong increase. Moreover,
this feature is amplified by the fact that the dominant tensor part of
DW weighs specifically the short range region by $r^{-3}$ (see
eq.~(\ref{t_dwba})). Comparing the approximations ``HM'' and ``DW''
one readily notices that at 1~MeV both give almost identical results
but with decreasing energies ``HM'' underestimates the Coulomb effects
more and more although the oscillations are quite similar except for
the size of the amplitude. 


\begin{figure}[ht]
\includegraphics[width=1.\columnwidth]{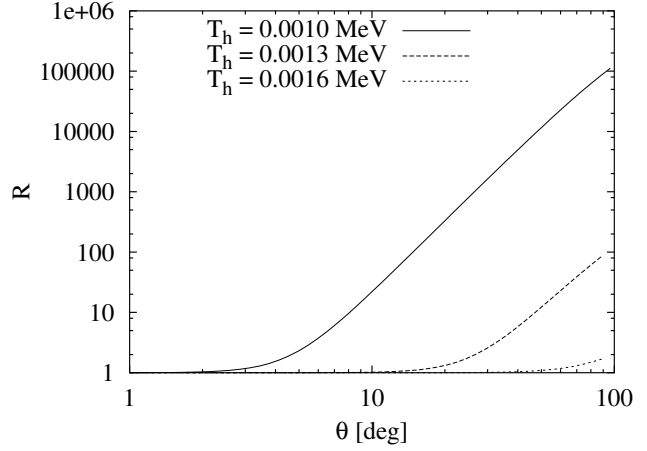}
\caption{Ratio of the unpolarized cross section with spin contribution
over the Rutherford cross section for three lab kinetic energies.}
\label{diffratio} 
\end{figure}

While for antiproton electron scattering the spin contribution to the
unpolarized differential cross section is strongly suppressed in
``DW'' with decreasing energy due to the increasing repulsive effect
of the Coulomb field, the opposite happens for proton electron
scattering. This is demonstrated in Fig.~\ref{diffratio} where the
ratio of the differential cross section in ``DW'' to the Rutherford
cross section is plotted. One readily notes that with increasing
scattering angle, corresponding to a decreasing impact parameter, for
which the influence of the Coulomb field increases, the ratio rapidly
becomes larger than one. However, the onset angle at which the ratio
starts to become larger than one increases rapidly with energy. It is
about 3$^\circ$ at $E=0.001$~MeV and already 60$^\circ$ at
$E=0.0016$~MeV. Since for the angle-integrated cross sections the
region of small angles is heavily weighted, the effect is still small
at $E=0.001$~MeV on the integrated cross section. But at smaller
energies the ``DW'' will fail. 

The relevant quantity for the build-up of the hadron polarization in a
storage ring is the total polarization transfer cross section which is
obtained by integrating the differential polarization transfer cross section
of (\ref{spintransfer_C}) over the solid angle up to a minimal
scattering angle which is determined by the requirement that the
impact parameter should not exceed a maximal value $b$, 
\beq
\theta_{min}=2\tan({\eta_c}/{l})
\eeq
with the classical angular momentum $l=pb$. For the total polarization
transfer cross section 
\beqa
\langle P_{ij}\sigma\rangle&=&\int d\Omega_h
P_{ij}\frac{d\sigma}{d\Omega_h} \nonumber\\
&=&\int_0^{2\pi} d\phi\int_{\theta_{min}}^\pi d\cos(\theta)
P_{ij}\frac{d\sigma}{d\Omega_h}\,,
\eeqa
the $\phi$-integration can be done analytically according to
(\ref{phi_xx}) through (\ref{phi_yz}). Obviously, the non-diagonal
components vanish and for the diagonal components one finds, using the
fact that the trace of $S^0_{ij}$ vanishes, i.e. 
\beq
S^0_{xx}+S^0_{yy}=-S^0_{zz}\,,
\eeq
the following results
\beqa
\langle P_{xx}\sigma\rangle&=&\langle P_{yy}\sigma\rangle\nonumber\\
&=&\frac{\pi c_s}{2p^2}\int_{\theta_{min}}^\pi 
\frac{d\cos(\theta)}{\sin^2(\theta/2)}
\Big(S_0^0- \frac{1}{2} S^0_{zz}\Big)\,,\\
\langle P_{zz}\sigma\rangle&=&\frac{\pi c_s}{2p^2}
\int_{\theta_{min}}^\pi 
\frac{d\cos(\theta)}{\sin^2(\theta/2)}
\Big(S_0^0+ S^0_{zz}\Big)\,.
\eeqa
Since, as already mentioned, the scalar contribution is negligible
compared to the tensor one, i.e.\ $S_0^0\ll S^0_{zz}$, one has the
simple relation
\beq
\langle P_{xx}\sigma\rangle=\langle P_{yy}\sigma\rangle
\approx-\frac{1}{2}\langle P_{zz}\sigma\rangle\,,
\eeq
and consequently it suffices to consider $\langle P_{zz}\sigma\rangle$
alone. 

\begin{figure}[ht]
\includegraphics[width=1.\columnwidth]{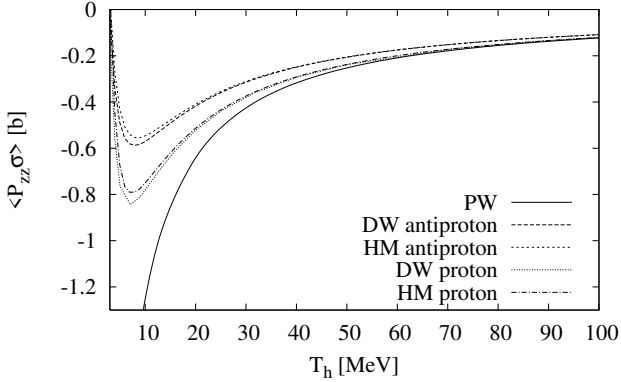}
\caption{The integrated polarisation transfer cross section 
$\langle P_{zz}\sigma\rangle$ for antiproton and proton electron
scattering in the c.m.\ frame as function of the hadron lab kinetic energy
for $b=10^{10}$~fm in PW, HM, and DW.}
\label{sigAP} 
\end{figure}

In Fig.~\ref{sigAP} we show first for the range of higher lab kinetic
energies between 3 and 100~MeV the result for the integrated polarization
transfer cross section $\langle P_{zz}\sigma\rangle$ for both,
antiproton and proton electron scattering for $b=10^{10}$~fm without,
i.e.\ in PW, and 
with inclusion of Coulomb effects for HM and DW approximations. While
near 100~MeV Coulomb effects are very small and all three
approximations give nearly the same result, the influence of the
Coulomb field becomes increasingly important with decreasing energy,
resulting in an increasing reduction of $\langle P_{zz}\sigma\rangle$
which is stronger for antiproton compared to proton electron scattering. In
this energy range HM and DW approximations give very similar results. 

\begin{figure}[ht]
\includegraphics[width=1.\columnwidth]{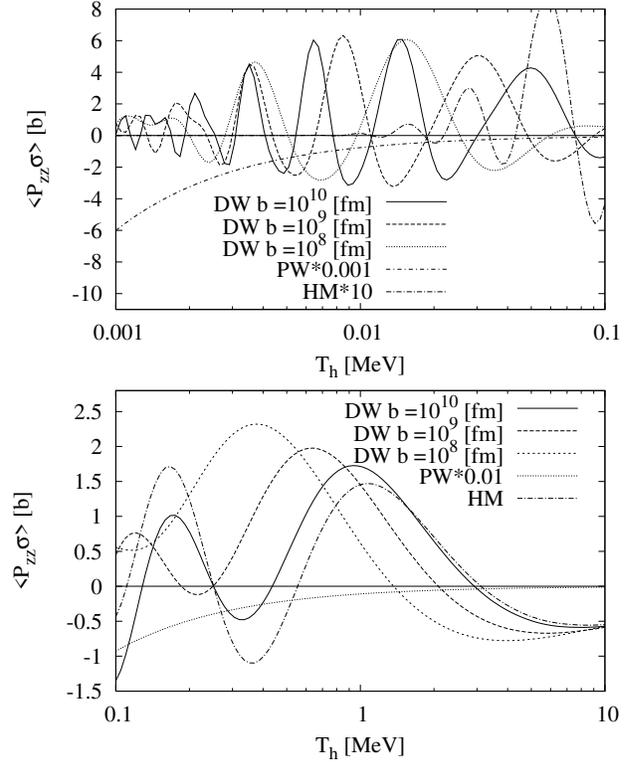}
\caption{The integrated polarisation transfer cross section 
$\langle P_{zz}\sigma\rangle$ for antiproton electron scattering
in the c.m.\ frame as function of the antiproton lab kinetic energy
for three values of the $b$-parameter. In the upper panel the results
in PW are divided by 1000 while those in HM are multiplied by 10. In
the lower panel the PW results are divided by 100.}
\label{sigA} 
\end{figure}

The lower energy range between 0.001 and 10~MeV is displayed in
Fig.~\ref{sigA} for antiproton electron scattering, again in PW, HM,
and DW approximations for $b=10^{10}$~fm. In order to study the 
dependence on the $b$-parameter, i.e.\ on the minimal scattering
angle, we show in addition results in DW for $b=10^{8}$ and
$10^{9}$~fm. It is 
apparent that below 1~MeV Coulomb effects continue to strongly
suppress $\langle P_{zz}\sigma\rangle$, but introduce rapid
oscillations with nearly constant amplitude except at very low
energies. Also the HM approximation shows such oscillations, but with 
shifted phase and rapidly decreasing amplitude leading to almost
vanishing cross sections below 0.01~MeV. The DW gives still
sizeable cross sections which essentially arise from the tensor part
because the scalar part is completely negligible. The decrease of the
$b$-parameter leads to a slight shift of the oscillations while the
amplitude is less affected. 

\begin{figure}[ht]
\includegraphics[width=1.\columnwidth]{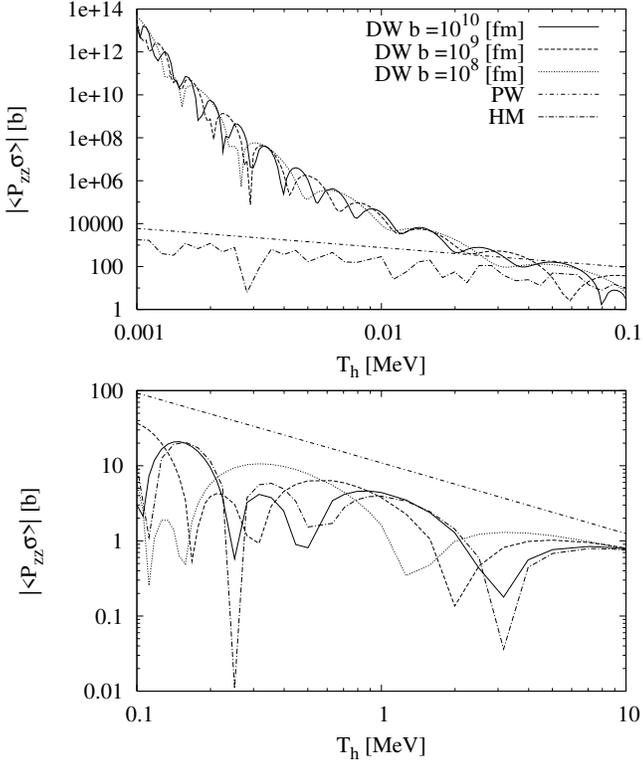}
\caption{Absolute value of the integrated polarisation transfer cross
section $|\langle P_{zz}\sigma\rangle|$ for proton electron scattering
in the c.m.\ function of the antiproton lab kinetic energy for three
values of the $b$-parameter.} 
\label{sigP} 
\end{figure}
\begin{figure}[ht]
\includegraphics[width=1.\columnwidth]{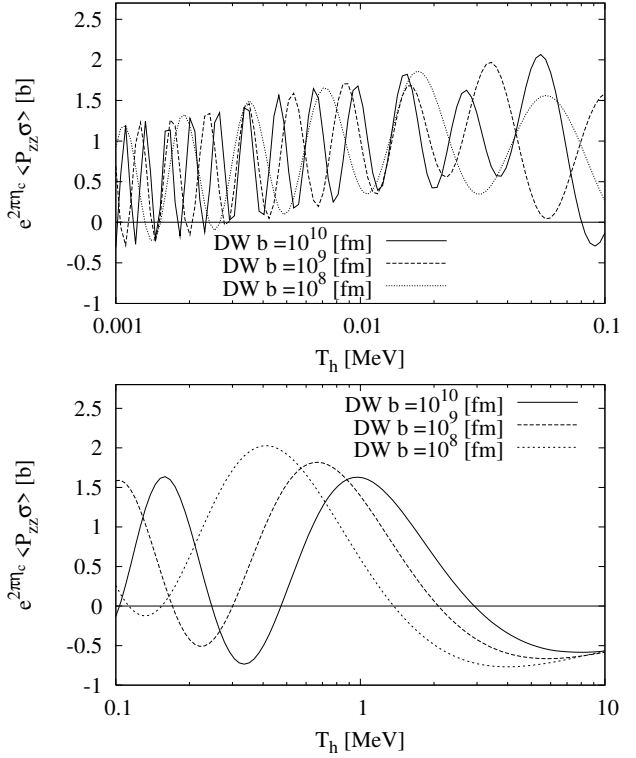}
\caption{The integrated polarisation transfer cross section 
$\langle P_{zz}\sigma\rangle$ for proton electron scattering
in the c.m.\ frame divided by the factor $e^{-2\pi\eta_c}$ as function
of the proton lab kinetic energy for three values of the
$b$-parameter.} 
\label{sigPnorm} 
\end{figure}

The corresponding results for proton electron scattering are shown in
Fig.~\ref{sigP}. In contrast to the antiproton case one notes here a
very rapid increase of the integrated polarization transfer cross sections with
decreasing energy. This rapid increase is expected from what has been
found for the differential polarization transfer cross section (see
Fig.~\ref{diffsigP}). As already said above, it is caused by the
strong attraction of the Coulomb field at small distances pulling in
the scattering wave towards the center. In order to make possible a
comparison with the 
antiproton case we display in Fig.~\ref{sigPnorm} again
$\langle P_{zz}\sigma \rangle$ with the essential difference that the
dominant factor $e^{-2\pi\eta_c}$ has been divided out. 
One notes a similar pattern as
for the antiproton case with the important exception that, if one takes
into account the normalization factor, the amplitude of the
oscillations increases very rapidly with decreasing energy so that in
the maxima $\langle P_{zz}\sigma\rangle$ becomes much larger than in
PW. 

\section{Summary and conclusions}                            \label{conc}
The polarization transfer in elastic hadron (proton and antiproton)
electron scattering from an initially polarized electron to the final
hadron has been calculated at low hadron kinetic energies with
inclusion of Coulomb effects in a distorted wave approximation. 

For antiproton electron scattering the influence of the repulsive
Coulomb field 
leads to a strong reduction of the polarization transfer cross section
compared to the plane wave approximation and in the angular dependence
of the differential cross section to an increasing oscillatory
behavior with decreasing energy. Consequently, the integrated
polarization transfer cross section is considerably smaller than for
the plane wave approximation, and exhibits oscillations with respect
to the energy dependence. In the maxima $\langle P_{zz}\sigma\rangle$ does
not exceed about 5~[b]. The position of the maxima depends slightly on
the minimal scattering angle determined by a maximal impact parameter.

Qualitatively the same oscillatory behaviour is found for proton
electron scattering when Coulomb effects are included, however, with
the essential difference of a 
rapidly increasing amplitude with decreasing energy, exceeding largely
the plane wave result. The reason for this is that the in this case
the attractive Coulomb field pulls the scattering wave towards smaller
distances, the effect of which increases strongly with decreasing
energy. This effect, however, limits the validity of the distorted wave
approximation towards very small energies, say below 1~keV. 

However, with respect to an experimental method to polarize antiprotons using
the polarization transfer from polarized positrons to antiprotons in a storage
ring, one has to include a possible initial hadron polarization and,
furthermore, to distinguish between polarization transfer without and with
spin-flip processes. This will be studied in a forthcoming
paper~\cite{Are07}. 

\section*{Appendix: The Coulomb integral}
For the evaluation of the integral
\beq
{D}_{ij}=\frac{1}{4\pi}\int d^3r 
\psi^{C(-)}_{\vec p^{\,\prime}}(\vec r)^*\frac{1}{r^3}
(3\hat{r}_i\,\hat{r}_j-\delta_{ij})\psi^{C(+)}_{\vec p}(\vec r)\,.
\eeq
one can 
make use of an integral representation of the
confluent hypergeometric function as recently proposed in~\cite{LeA01}
(for convenience we set here and in the following $\eta_c=\eta$)
\beqa
_1F_1(-i\eta,1;ix)&=& Q(\eta)\int_0^1 dt \,f(t,\eta)^*\,e^{ixt}\nonumber\\
&&\times(1-(1-t)ix)\,,
\eeqa
with
\beq
Q(\eta)=\frac{\sinh{\pi\eta}}{\pi\eta}\,\,\mbox{and}\,\, f(t,\eta)=
e^{i\eta\ln{\frac{t}{1-t}}}.
\eeq
Substituting $t$ by $1-t$, one finds the equivalent form
\beqa
_1F_1(-i\eta,1;ix)&=& Q(\eta)\int_0^1 dt \,f(t,\eta)\,e^{ix(1-t)}
\,(1-ixt)\nonumber\\
&&\hspace*{-1cm}= Q(\eta)\int_0^1 dt\,f(t,\eta)\,(1+{\cal O}_t)\,e^{ix(1-t)}\,.
\label{intrep}
\eeqa
where ${\cal O}_t=t\frac{\partial}{\partial t}$.
Using the relation between the cartesian and the spherical form of a
symmetric, traceless tensor of second rank
\beq
3\hat{r}_i\,\hat{r}_j-\delta_{ij}=\sum_M c_{ij,M} Y_{2M}(\hat{r})\,,
\eeq
where we do not need to specify the coefficients $c_{ij,M}$, 
one can write
\beq
{D}_{ij}=\sum_M c_{ij,M} D_{2M}
\eeq
with
\beqa
D_{2M}&=&\frac{N(\eta)^2}{4\pi}\int \frac{d^3r}{r^3}\,
_1F_1(-i\eta,1;i(pr+\vec p^{\,\prime}\cdot\vec r\,))\,\nonumber\\&&
\hspace*{-1cm}\times Y_{2M}(\hat{r})\,
e^{i(\vec p-\vec p^{\,\prime}\,)\cdot\vec r}\,
_1F_1(-i\eta,1;i(pr-\vec p\cdot\vec r\,))\,.\label{d2M}
\eeqa
Inserting now the integral representation of (\ref{intrep}), one finds
\beqa
D_{2M}&=&\frac{(N(\eta)Q(\eta))^2}{4\pi}\int_0^1 dt\,f(t,\eta)\,(1+{\cal
O}_t)\,\nonumber\\
&&\int_0^1 dt'\, f(t',\eta)\,(1+{\cal O}_{t'})\,I_{2M}(t,t')
\eeqa
 with
\beq
I_{2M}(t,t'\,)=\int \frac{d^3r}{r^3}\,Y_{2M}(\hat{r})\,e^{ipr(2-t-t')}\,
e^{i(\vec p\,t-\vec p^{\,\prime}\,t')\cdot\vec r}\,.
\eeq
The integration over $\vec r$ can be done analytically yielding first
for the angular integration
\beqa
I_{2M}(t,t'\,)&=&-4\pi\,Y_{2M}(\hat{a}(t,t'))\nonumber\\&&
\int_0^\infty\frac{dr}{r}\,e^{ipr(2-t-t')}\,j_2(a(t,t')r)
\,.
\eeqa
By a transformation of the integration variable, one obtains
\beqa
I_{2M}(t,t'\,)&=&-4\pi\,Y_{2M}(\hat{a}(t,t'))
\int_0^\infty\frac{dx}{x}\,e^{ic(t,t')x}\,j_2(x)\nonumber\\
\nonumber\\
&=&-4\pi\,Y_{2M}(\hat{a}(t,t'))I(c(t,t'))\,,
\eeqa
where we have introduced
\beqa
\vec a(t,t')&=&\vec p\,t-\vec p^{\,\prime}\,t'\,,\\
a(t,t')&=&|\vec a(t,t')|=p\,g(t,t')\,,\\
g(t,t')&=&[t^2+t'^2-2tt'\cos{\theta}]^{1/2}\,,\\
c(t,t')&=&\frac{2-t-t'}{g(t,t')}\,,
\eeqa
and $\theta$ denotes the scattering angle in the c.m.\ frame. This
integral is solved analytically
\beqa
I(c)&=& \int_0^\infty\frac{dx}{x}\,e^{icx}\,j_2(x)\nonumber\\
&=&\frac{1}{3}
-\frac{1}{2}\,c^2\nonumber\\&&-\frac{1}{4}\,c(1-c^2)\,
\Big(\ln\Big|\frac{c+1}{c-1}\Big|-i\pi\,\Theta(1-c)\Big)
\,.
\eeqa
With the help of 
\beqa
A_{ij}(t,t')&=&3\hat{a}_i(t,t')\,\hat{a}_j(t,t')-\delta_{ij}\nonumber\\
&=&\sum_M c_{ij,M}\,Y_{2M}(\hat{a}(t,t'))\,,
\eeqa
the following form for the tensor in (\ref{dij}) is obtained
\beqa
{D}_{ij}&=&-N(\eta)^2Q(\eta)^2\int_0^1 dt\,f(t,\eta)\,
(1+{\cal O}_t)\,\nonumber\\&&\hspace*{-.5cm}\int_0^1 dt'\, 
f(t',\eta)\,(1+{\cal O}_{t'})\,A_{ij}(t,t')\,I(c(t,t'))\,.
\eeqa

The remaining integrations over $t$ and $t'$ will be done
numerically. To this end one would have to evaluate the derivatives
\beqa
(1&+&{\cal O}_t)\,(1+{\cal O}_{t'})\,A_{ij}(t,t')\,I(c(t,t'))=\nonumber\\
&&\hspace*{4cm}A_{ij}(t,t')\,I(c(t,t'))\nonumber\\&&+
(t\frac{\partial }{\partial t}+t'\frac{\partial }{\partial t'}
+tt'\frac{\partial ^2}{\partial t\partial t'})
A_{ij}(t,t')I(c(t,t'))\,.
\eeqa
Another possibility is to eliminate them by partial integration, i.e.\
\beqa
\int_0^1 dt f(t',\eta)\,(1+t'\frac{d}{dt'})g(t')&=&\lim_{t\to 1-}
\Big[t'f(t',\eta)g(t')\Big|_0^t\nonumber\\
&&\hspace*{-1.5cm}-\int_0^t dt't'g(t')\frac{d}{dt'}f(t',\eta)\Big]\,,
\eeqa
and for 
\beq
f(t',\eta)=e^{i\eta\ln{\frac{t'}{1-t'}}}\,,
\eeq
one finds
\beqa
\int_0^1 dt f(t',\eta)\,(1+t'\frac{d}{dt'})g(t')&=&\lim_{t\to 1-}
\Big[tf(t,\eta)g(t)\nonumber\\
&&\hspace*{-1.8cm}-i\eta\int_0^t \frac{dt'}{1-t'}\,g(t')f(t',\eta)\Big].
\label{partialint}
\eeqa
That the limit $t\to 1-$ exists can be seen by using the following 
identity (for $t< 1$)
\beq
e^{-i\eta\ln{(1-t)}}=1+i\eta\int_0^t
\frac{dt'}{1-t'}\,e^{-i\eta\ln{(1-t')}}\,. 
\eeq
Thus (\ref{partialint}) becomes
\beqa
\label{partialinta}
&&\int_0^1 dt f(t',\eta)\,(1+t'\frac{d}{dt'})g(t')=
g(1)\\&&\hspace*{1cm}+i\eta\int_0^1\frac{dt'}{1-t'}\,e^{-i\eta\ln{(1-t')}}
\Big(g(1)-e^{i\eta\ln{t'}}g(t')\Big)\,.\nonumber
\eeqa
In particular, one obtains with $g(t)=e^{ixt}$ an integral representation of
the confluent hypergeometric function equivalent to (\ref{intrep}), namely
\beqa
&&_1F_1(-i\eta,1;ix)= \\
&&Q(\eta)\Big[1+i\eta\int_0^1\frac{dt'}{1-t'}\,
e^{-i\eta\ln{(1-t')}}
\Big(1-e^{i\eta\ln{t'}}e^{ixt'}\Big)\Big]\,.\nonumber
\eeqa

Using the representation (\ref{partialinta}) for both integrations over $t$
and $t'$, one obtains as final form for the tensor 
\beqa
{D}_{ij}&=&-N(\eta)^2Q(\eta)^2
\Big[\widetilde{D}_{ij}(1)\\&&\hspace*{-1.5cm} 
+i\eta\int_0^1\frac{dt}{1-t}\,e^{-i\eta\ln{(1-t)}}
\Big(\widetilde{D}_{ij}(1)-e^{i\eta\ln{t}}
\widetilde{D}_{ij}(t)\Big)\Big]\,,\nonumber
\eeqa
where
\beqa
\widetilde{D}_{ij}(t)&=&A_{ij}(t,1)\,I(c(t,1))\\&&\hspace*{-1cm}+
i\eta\int_0^1\frac{dt'}{1-t'}\,e^{-i\eta\ln{(1-t')}}
\Big(A_{ij}(t,1)\,I(c(t,1))\nonumber\\&&
-e^{i\eta\ln{t'}}A_{ij}(t,t')\,I(c(t,t'))\Big)\,.\nonumber
\eeqa
For $\eta=0$, which means neglecting Coulomb effects, one finds
with $N(0)=1$ and $Q(0)=1$
\beqa
D_{ij}|_{\eta=0}&=&\widetilde{D}_{ij}(1)|_{\eta=0}\nonumber\\
&=&-A_{ij}(1,1)I(c(1,1))\,,
\eeqa
and furthermore with $c(1,1)=0$, thus $I(c(1,1))=1/3$ and $\vec
a(1,1)=\vec q$ finally
\beq
D_{ij}|_{\eta=0}=-(\hat q_i\hat q_j-\frac{1}{3}\delta_{ij})\,.
\eeq
Thus for this case ($\eta=0$) eqs.~(\ref{s_dwba}) and (\ref{t_dwba})
indeed reduce to (\ref{s_born}) and (\ref{t_born}), respectively. 

With respect to the dependence of ${D}_{ij}$ on the sign of $\eta$ it
is useful to separate the normalization factor $N(\eta)^2$ and
to split ${D}_{ij}$ into two contributions according to the real and
imaginary part of $I(c)$, indicating the $\eta$-dependence only, 
\beq
{D}_{ij}(\eta)=N(\eta)^2Q(\eta)^2
(R_{ij}(\eta)+I_{ij}(\eta))\label{sym}\,,
\eeq
where
\beqa
R_{ij}(\eta)&=&-\int_0^1 dt\,f(t,\eta)\,
(1+{\cal O}_t)\,\int_0^1 dt'\,f(t',\eta)\,(1+{\cal
O}_{t'})\nonumber\\&&\times A_{ij}(t,t')\,\Re eI(c(t,t'))\,,\\ 
I_{ij}(\eta)&=&-i\int_0^1 dt\,f(t,\eta)\,
(1+{\cal O}_t)\,\int_0^1 dt'\,f(t',\eta)\,(1+{\cal
O}_{t'})\nonumber\\&&\times A_{ij}(t,t')\,\Im mI(c(t,t'))\,.
\eeqa
Both contributions are complex. They possess the following simple
symmetry under a sign change of $\eta$
\beqa
R_{ij}(-\eta)&=&(R_{ij}(\eta))^*\,,\label{symreal}\\
I_{ij}(-\eta)&=&-(I_{ij}(\eta))^*\,,\label{symimag}
\eeqa
because one has
\beq
f(t,-\eta)=(f(t,\eta))^*\,.
\eeq
Thus it suffices to calculate $R_{ij}$ and $I_{ij}$ for one sign of
$\eta$. 

The numerical evaluation of the integrals is straight\-for\-ward. However,
it turns out, that for the practical evaluation a transformation of
the integration variable is advantageous by setting $\tau=1/(1-t)$
resulting in
\beqa
\int_0^1\frac{dt}{1-t}\,e^{-i\eta\ln{(1-t)}}(g(1)-g(t))&=&\\
\int_1^\infty\frac{d\tau}{\tau}\,e^{i\eta\ln\tau}(g(1)&-&g(t(\tau)))\,,
\nonumber 
\eeqa
which shows good convergence for $\tau \to \infty$.

\begin{acknowledgement}
I would like to thank Thomas Walcher for the motivation for this
work and, furthermore, him, Erwin Alt, and Michael Schwamb for many useful
discussions. 
This work has been supported by the SFB~443 of the Deutsche
For\-schungs\-gemeinschaft (DFG).
\end{acknowledgement}

\end{document}